# Origin of low sodium capacity in graphite and generally weak substrate binding of Na and Mg among alkali and alkaline-earth metals

Yuanyue Liu,[*,1,2] Boris Merinov[1] and William A. Goddard III[*,1]

[1]*Materials and Process Simulation Center*
[2]*the Resnick Sustainability Institute*
*California Institute of Technology, Pasadena, CA 91125, USA*

[*]Correspondence to: yuanyue.liu.microman@gmail.com and wag@wag.caltech.edu

**Abstract:** It is well known that graphite has a low capacity for Na but a high capacity for other alkali metals. The growing interest in alternative cation batteries beyond Li makes it particularly important to elucidate the origin of this behavior, which is not well understood. In examining this question, we find a quite general phenomenon: among the alkali and alkaline earth metals, *Na and Mg generally have the weakest chemical binding to a given substrate*, compared with the other elements in the same column of the periodic table. We demonstrate this with quantum mechanics calculations for a wide range of substrate materials (not limited to C) covering a variety of structures and chemical compositions. The phenomenon arises from the competition between trends in the ionization energy and the ion-substrate coupling, down the columns of the periodic table. Consequently, the cathodic voltage for Na and Mg is expected to be lower than those for other metals in the same column. This generality provides a basis for analyzing the binding of alkali and alkaline-earth metal atoms over a broad range of systems.

**Significance:** The growing demand for energy storage urges the development of alternative cation batteries, which calls for a systematic understanding of binding energetics. We discover a general phenomenon for binding of alkali and alkali-earth metal atoms with substrates, which is explained in a unified picture of chemical bonding. This allows us to solve the long-standing puzzle of low Na capacity in graphite, predict the trends of battery voltages, and also forms a basis for analyzing the binding of alkali and alkaline-earth metal atoms over a broad range of systems.

\body

**Main text**: Development of alternative cation batteries beyond Li could solve issues related with cost, stability, and other performance characteristics(1-3). Na is an obvious candidate, but its storage in graphite (the commercial anode for Li-ion battery) is rather poor, with an electrochemical capacity < ~35 mAh/g(1, 4-7). Surprisingly, other alkali metals have a high capacity (~ hundreds) in graphite(8). In order to form a basis for improving the battery performance, we seek to understand the anomalously low capacity for Na.

One explanation in the literature for the low Na capacity is: Na intercalation expands graphite from its favorable interlayer spacing, by a greater amount than Li, leading to a higher strain energy for graphite and, therefore, a less favorable formation energy for Na-graphite compound compared to the Li analog(9-11). However, this explanation would suggest that graphite should have a low capacity for K, Rb and Cs because of their even larger size, which is in stark contrast to the experimentally observed capacity dramatically higher than that for Na. This inconsistency calls for a revisit of the origin of the low Na capacity in graphite.

In this work, we use Quantum Mechanical methods to show that the Na anomaly has its roots in to a general phenomenon: among alkali metals (denoted as M) and alkaline-earth metals (denoted as EM), Na and Mg generally have the weakest binding to a given substrate, independent of variations in substrate structure and chemistry! This phenomenon results from the competition between the ionization of the metal atom and the ion-substrate coupling, which have opposite trends along the columns. The universality of this phenomenon provides the basis for analyzing trends in binding of alkali and alkaline-earth metals over a broad range of systems, and offers guidance for designing improved systems.

Our Density functional theory (DFT) calculations used the Vienna Ab-initio Simulation Package (VASP)(12, 13) with projector augmented wave (PAW) pseudopotentials (14, 15). The Perdew-Burke-Ernzerhof (PBE) exchange-correlation functional(16) including van der Waals corrections (DFT-D3)(17) is employed to model the graphite and its compounds, while in other cases where the van der Waals interaction is insignificant, we use PBE only. The plane-wave cut-off energy is 400 eV, with sufficient Monkhorst-Pack sampled k-points(18) (for example 15x15x7 for graphite). All structures are fully relaxed until the final force on each atom becomes less than 0.01 eV/Å. We used a single M or EM atom with 6x6 unit cells of each substrate material to model the binding with graphene, its derivatives, $MoS_2$, $SnS_2$, and $TiS_2$, while a 4x4 cell is applied for $V_2O_5$. For Pt(111), we used a slab consisting of 6x6 surface cells, and three layers, with the bottom layer fixed in the same plane.

We consider first the case of graphite. Fig. 1 shows the calculated formation energy ($E_f$) of M-graphite compounds, where M = Li, Na, K, Rb, and Cs. The $E_f$ is defined as:

$E_f = [\ E(\text{tot}) - n_M\ E(\text{bulk M}) - n_C\ E(\text{graphite})\ ] / (n_M + n_C)$   (1)

where $E(\text{tot})$ is the total energy of compound, $E(\text{bulk M})$ is the energy per M atom for the bulk metal, $E(\text{graphite})$ is the energy of C in graphite, $n_M$ and $n_C$ are the number of M and C atoms in the compound. Here we focus on $MC_6$ and $MC_8$ since these stoichiometries are commonly found in non-Na compounds. We find that, the $E_f$ follows the order: Na > Li > K > Rb > Cs, where all Ms except Na have negative $E_f$ with graphite, a result consistent with calculations in the literature which use different method(19). Thus Na-graphite compounds with high Na contents are not thermodynamically stable, in contrast to the other four M-graphite compounds, in agreement with the experimentally observed low-Na capacity.

To understand why the $E_f$, has a maximum at Na, we partition the formation process for M-graphite compounds into three steps, as illustrated in Fig. 2a. First, the bulk metal is evaporated to form isolated atoms, with an energy cost of $E_d$ (i.e., the cohesive energy). Second, the graphite crystal is strained to the configuration identical to that of the M-intercalated graphite, with an energy cost of $E_s$. This straining includes both interlayer expansion and in-plane stretching. Third the M atoms are intercalated into the strained graphite, with an energy drop by $E_b$ (to be consistent with the other terms, we define $E_{-b} = -E_b$; a stronger binding corresponds a lower $E_{-b}$). According to Hess's law:

$E_f = E_d + E_s + E_{-b}$      (2)

This analysis helps identify the dominating contribution. Although some terms are difficult to measure experimentally (such as $E_s$ and $E_b$), all are calculated easily using DFT.

As shown in Fig. 2b, neither $E_d$, $E_s$, nor their combination has a trend similar to $E_f$, which suggests that they do not embody the origin of the low Na capacity. For both $MC_8$ and $MC_8$, $E_s$ increases monotonically as M moves down the periodic table due to the increasing size of M atoms, while $E_d$ decreases as a result of weakened cohesion. The combination of $E_s$ and $E_d$ shows a monotonic drop, indicating that the change of $E_d$ overwhelms that of $E_s$. Nevertheless, none of them explain the maximum of $E_f$ at Na.

On the other hand, $E_{-b}$ exhibits a maximum at Na, similar to that of $E_f$. This suggests that the low Na capacity is related directly to $E_{-b}$. In particular, compared with Li, the Na binding is so weak that it exceeds the decrease of $E_s + E_d$, making $E_f$ higher for Na.

This weaker binding of Na compared with Li has been reported for other intercalation compounds, which has been proposed to account for the observed lower cathodic voltage(20). Here we find that, of all the 5 alkali metals, Na always has the weakest binding for any given substrate. We first examine dilute M adsorption on graphene (Fig. 3a), which shows a maximum of $E_{-b}$ at Na. Then we modify the adsorption sites by incorporating structural defects or foreign atoms. Remarkably, Na always leads to the weakest binding (Fig. 3a). We continue the test by considering other two-dimensional non-C materials that have been tested for batteries, namely $MoS_2$, $TiS_2$, $SnS_2$, and $V_2O_5$ (Fig. 3b). We then extended this test to the surface of a typical bulk materials Pt(111) (Fig. 3b; as commonly found in Pt-based catalysis in alkaline solution). In all cases, Na has the weakest binding among all five alkali metals, independent of the detailed substrate chemistry/structure. This general phenomenon calls for a unified explanation.

$E_{-b}$ is the energy change when an M atom moves from the vacuum to the binding site of the substrate. We consider this process to first involve ionization of M by transferring its charge to the substrate, with an energy change by $E_{ion}$. This is followed by the coupling of the cation to the substrate (negatively charged) with an energy decrease of $E_{cp}$ (which includes the electrostatic and other quantum-mechanical interactions), as illustrated in Fig. 4a. Therefore,

$$E_{-b} = E_{ion} + E_{cp} \qquad (3)$$

As M moves down the periodic table, the ionization potential decreases: 5.4 (Li), 5.1 (Na), 4.3 (K), 4.2 (Rb), and 3.9 (Cs), which favors the binding and results in a decrease of $E_{ion}$. Note that there is an abrupt drop in the ionization potential between Na and K. However, at the same time, the distance between cation and the substrate becomes larger, which weakens their coupling leading to an increase of $E_{cp}$. To quantify this competition, Fig. 4b shows the $E_{ion}$ and $E_{cp}$ with respect to those of Li on the same substrate. The relative $E_{ion}$ is approximated by the difference in the atomic ionization potential (IP), and the relative $E_{cp}$ is then derived from Eq. (3). Since the $E_{cp}$ increases smoothly from Li to Cs while the $E_{ion}$ drops dramatically from Na to K (in other words, the coupling strength increases smoothly from Cs to Li while the ionization cost drops dramatically from K to Na), we obtain maximum of $E_{-b}$ for Na.

The $E_{cp}$ is a general term which includes various kinds of interactions that are difficult to separate. However, for the cases of M adsorption on metal (e.g. graphene and Pt (111)), one could expect the difference in $E_{cp}$ of different Ms is dominated by the electrostatic contribution, which is ~ $-14.38/(2*d)$, where $d$ is the distance between the cation and the substrate according to the image charge method. Therefore in these cases:

$$\Delta E_{-b} = \Delta E_{ion} + \Delta E_{cp} \sim \Delta[-14.38/(2*d)] + \Delta IP \qquad (4)$$

As shown in Figure 4c, the trend of $E_{-b}$ given by equation is similar to that calculated by using DFT, which validates that competition between $E_{ion}$ and $E_{cp}$ is reason for a maximum of $E_{-b}$ at Na. Indeed, this weak Na binding is also found in diatomic molecules M-X (where X=F, Cl, Br, I and OH)(21), which can be explained similarly.

It is interesting to consider whether the non-monotonic trend of the $E_{-b}$ is present in other columns. Based on the above explanations, we anticipate that this can be observed in the columns where $E_{ion}$ and $E_{cp}$ have a reverse trend when moving down the periodic table. The $E_{cp}$ perhaps always increases as the atomic size gets larger; however, $E_{ion}$ does not always decrease, given the fact that the IP or electron affinity is non-monotonic for most groups. In fact, only the first two groups have a notably decreasing IP. Therefore, one may expect a similar phenomenon occurs for the EM group. Indeed, our calculations show that among the EM elements, Mg generally has the weakest binding (Fig. 5) with a given substrate. This is consistent with the experimental fact that Mg has a low capacity in graphite, similar to the case of Na (8, 22). Note that Be and Mg are only physisorbed on pristine or nitrogen-doped graphene with the $E_{-b}$ ~ -30 meV/EM, significantly weaker than other EMs. This is because the work function of pristine or nitrogen-doped graphene is too low to allow for a charge transfer from Be/Mg, as shown by the band structures in the SI. These systems are perhaps not practically interesting as the physisorbed adatoms could easily detach or cluster. It should also be noted that, Be tends to have a stronger covalence than other EMs, due to its high IP and small size. This might be the reason for the significantly enhanced binding of Be with the mono-vacancy in graphene, in which case the $E_{cp}$ contributes a large energy drop.

For cathode materials, the weak binding with the metal atoms results in a low cathodic voltage. Therefore we anticipate that Na and Mg have a low cathodic voltage compared with other metals in the same columns. Indeed, the Na case has been verified by explicitly calculating the cathodic voltage for various intercalation compounds(20). On the other hand, a weak binding with anode is desired to enhance the voltage when connected with a cathode. However, this usually sacrifices capacity, as seen in the case of Na in graphite. To improve the Na capacity, it is necessary to reduce $E_f$. This can be achieved by using a pre-strained graphite (i.e. reducing the $E_s$ term in Eq. 2) with expanded interlayer spacing through intercalation of some other species, as has been demonstrated experimentally(9). Our calculations show that the optimum $E_f$ for Na-graphite is reached when the graphite interlayer distance is expanded to 4.3 Å, providing a target value for experimental design. Indeed such an expanded graphite using organic pillars has been shown computationally to provide very promising hydrogen storage of 6.5 wt% at room temperature, meeting the DOE requirement(23). Alternatively, the $E_f$ might be reduced by enhancing the binding, which might be achieved through incorporating defects(24, 25).

In summary, we use Quantum Mechanical calculations, to find a general phenomenon— among alkali and alkaline-earth metals, Na and Mg generally have the weakest binding for a given substrate. We show that this results from the competition between the ionization of the metal atom, and the ion-substrate coupling. This finding elucidates the origin of low Na capacity in graphite, predicts the voltage trends for alkali and alkaline-earth metal ion batteries, and provides a basis for analyzing the binding of alkali and alkaline-earth metal atoms in a broad range of systems.

**Acknowledgements**:

Y.L. thanks Drs. Brandon Wood, Suhuai Wei, and Jiayu Wan for helpful discussions and Brandon Wood for providing access to the Lawrence Livermore National Laboratory computational resources, which were used for some of the computations (supported under the Laboratory Directed Research and Development Program). Most of the calculations were performed on National Energy Research Scientific Computing Center, a Department of Energy (DOE) Office of Science User Facility supported by the Office of Science of the US DOE under Contract DE-AC02-05CH11231. Y.L. acknowledges the support from Resnick Prize Postdoctoral Fellowship at Caltech. This research was funded by the Bosch Energy Research Network and by NSF (CBET 1512759).

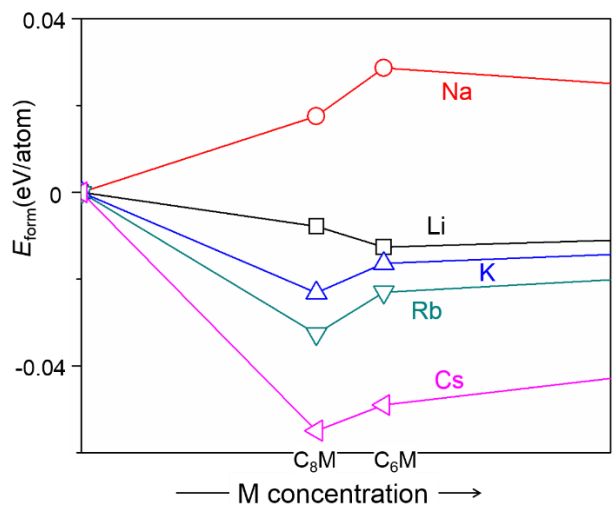

**Fig. 1**. Calculated formation energies (Eq. 1) of alkali metal (M)-graphite compounds. Note that in contrast to other Ms, $NaC_6$ and $NaC_8$ have a positive formation energy.

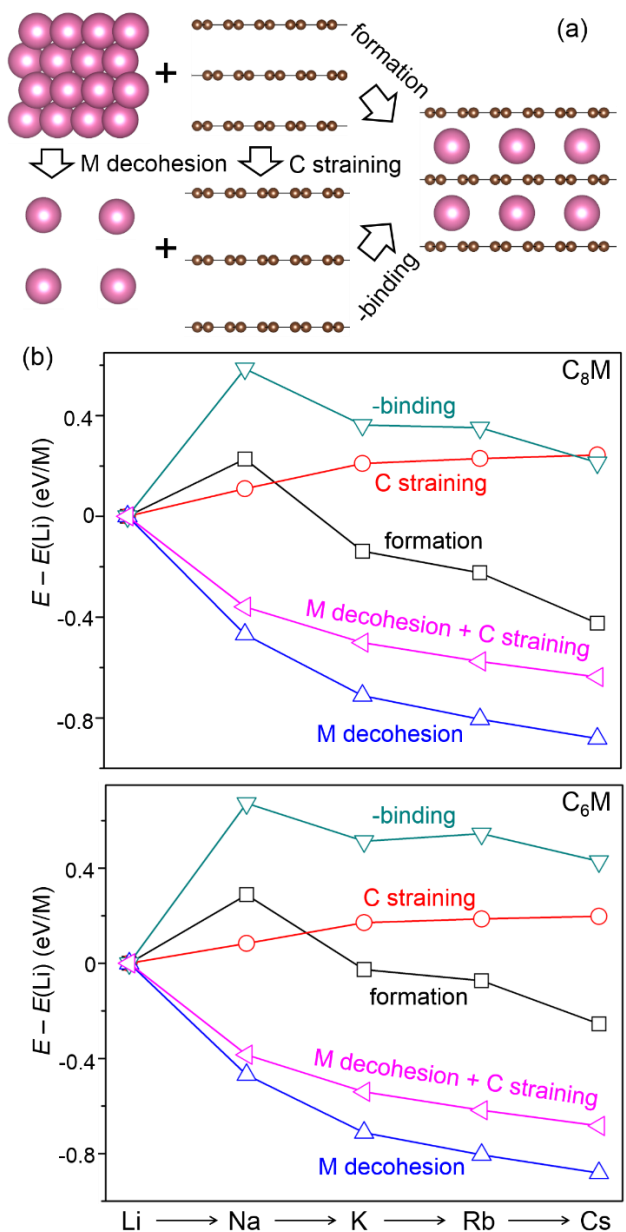

**Fig. 2**. (a) Partition of the formation process of M-graphite compound into separate steps. M: pink; C: brown. (b) The energetics of each step, relative to those of Li (see Eq. 2 and the related text). '-binding' means the reverse of binding energy, i.e. $E_{-b}$ in Eq. 2. Note that only $E_{-b}$ shows a trend similar with $E_f$.

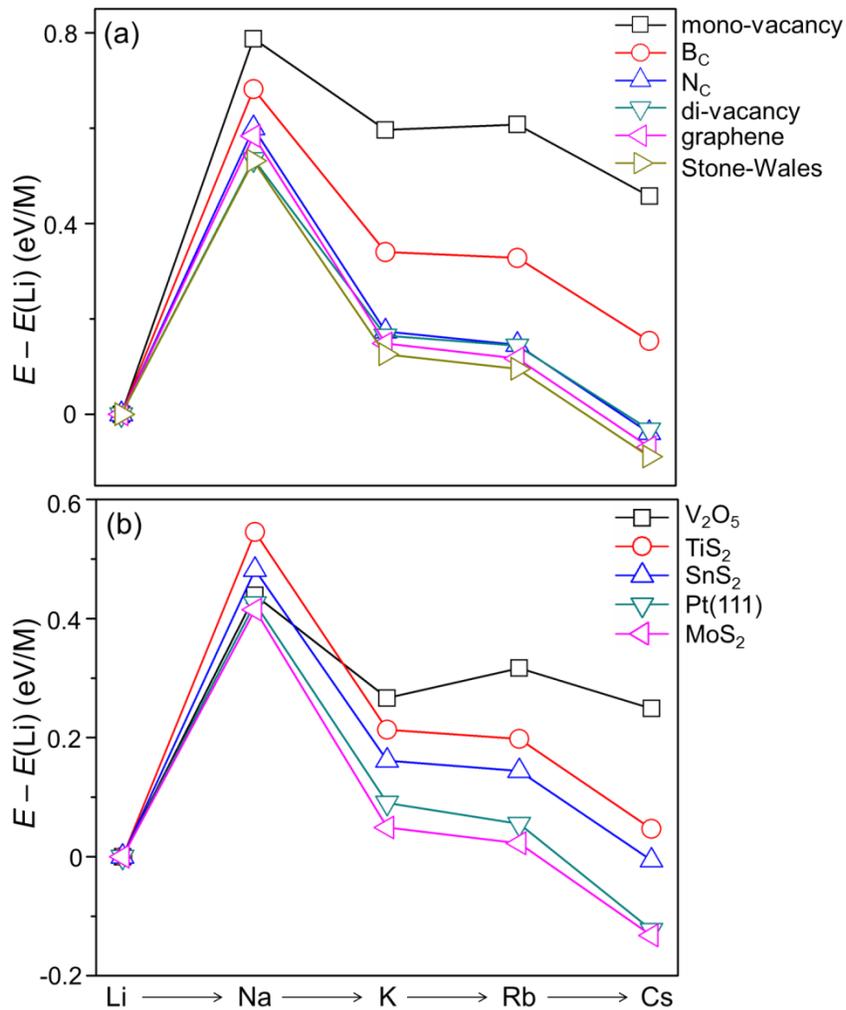

**Fig. 3**. $E_{-b}$ (the negative of the binding energy) for alkali metals binding to various substrate materials, relative to that of Li (the absolute values can be found in the SI). (a) graphene and its derivatives (with defects or substitutional foreign atoms); (b) non-C materials. Note that Na always has the weakest binding among alkali metals.

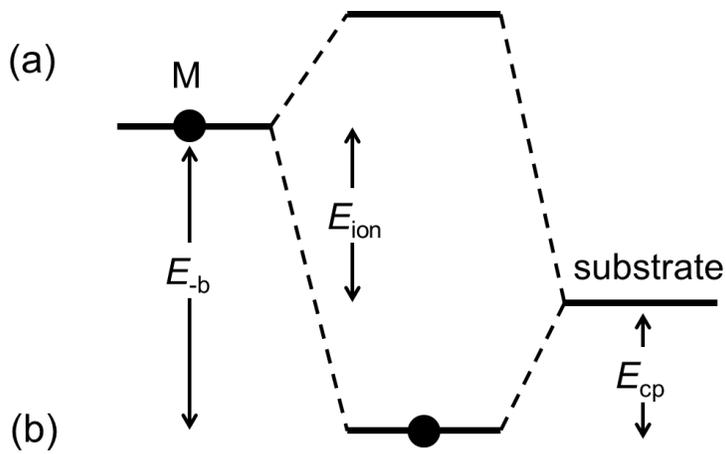
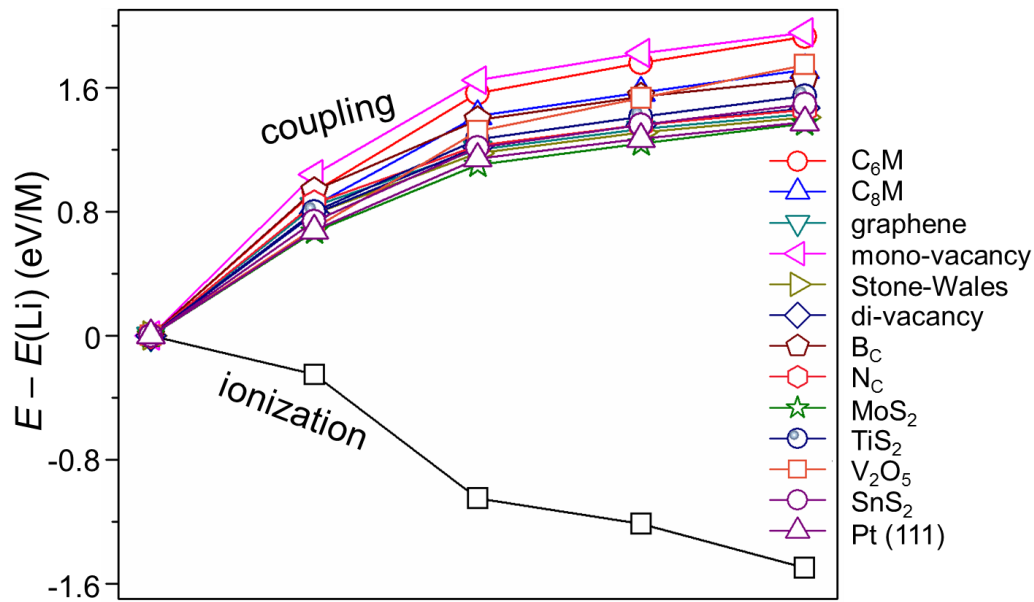
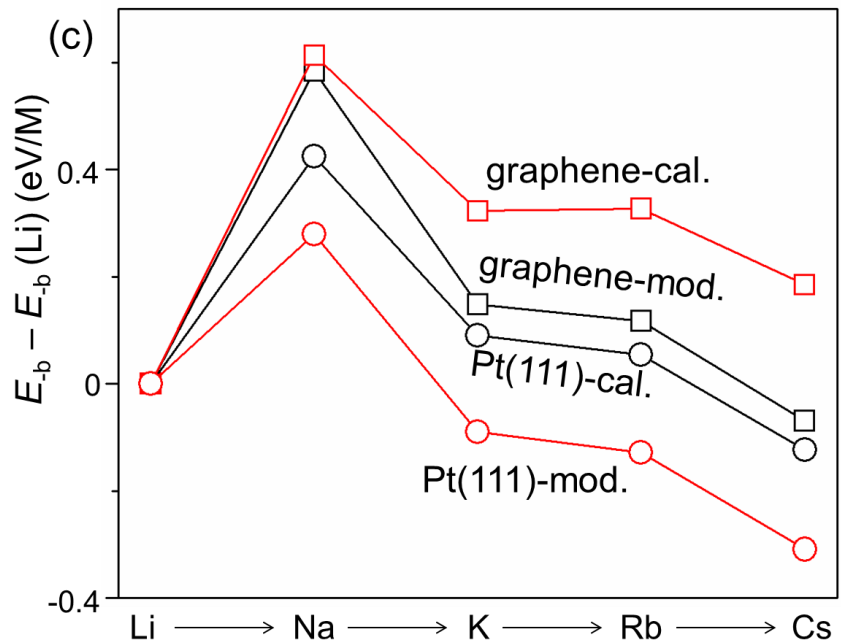

**Fig. 4**. (a) Schematic illustration of the binding between the alkali metal (M) and the substrate. (b) Evolution of the ionization energy ($E_{ion}$) and the coupling energy ($E_{cp}$) as M moves down in the periodic table. (c) Comparison of the calculated $E_{-b}$ (relative to that of Li) with that based on the model of Eq. 4.

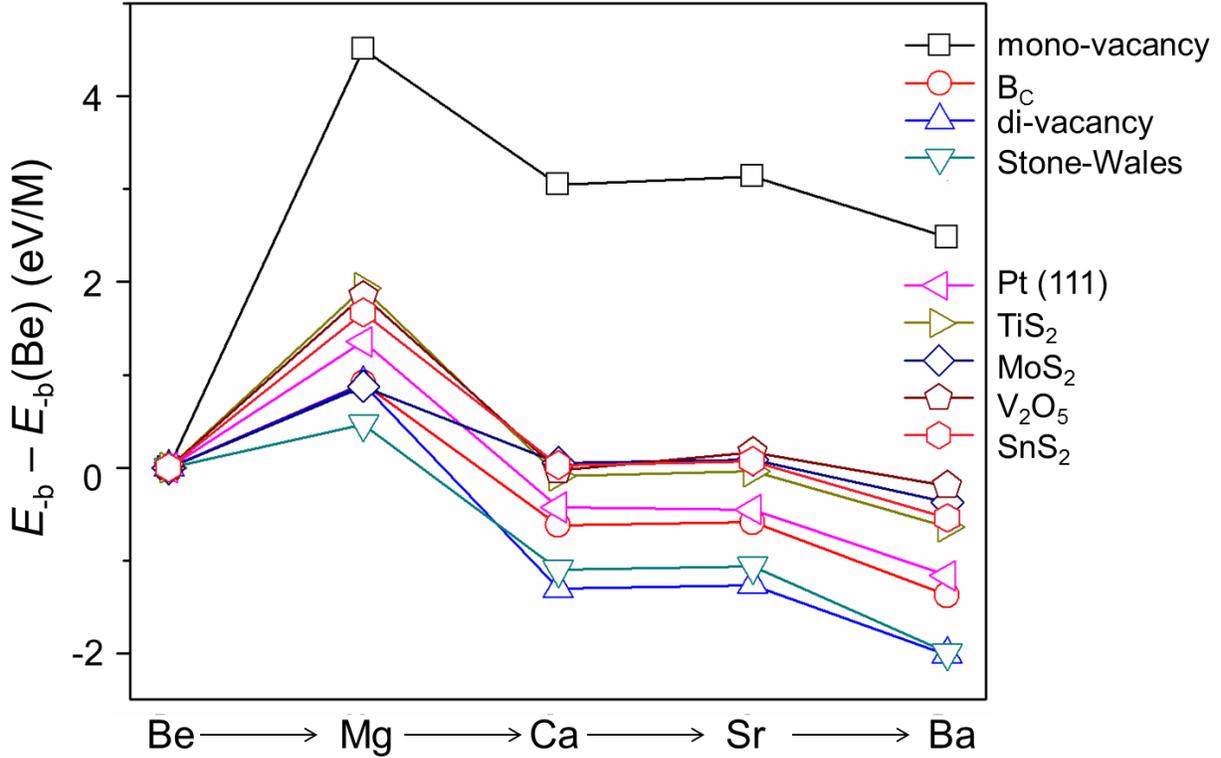

**Fig. 5**. $E_{-b}$ (the negative of the binding energy) for alkaline-earth metals binding to various substrate materials, relative to that of Be (the absolute values can be found in the SI). Note that Mg always has the weakest binding.